\documentclass[a4paper,11pt]{scrartcl}
\usepackage[british]{babel}
\usepackage[utf8]{inputenc}
\usepackage{amsthm}
\usepackage{amssymb}
\usepackage{amsmath}
\usepackage{nicefrac}
\usepackage{float}
\usepackage{setspace}
\usepackage{url}
\usepackage{bm}
\usepackage{mathtools}
\usepackage{makecell}
\usepackage{caption}
\usepackage{subcaption}
\usepackage{relsize}
\usepackage{enumerate}
\usepackage{listings}
\usepackage{tabu}
\usepackage{color}
\usepackage{cprotect}
\usepackage{graphicx}
\usepackage{makeidx}
\usepackage{dsfont}
\usepackage[ruled]{algorithm2e}
\usepackage{lipsum} 
\usepackage{subfiles} 
\usepackage[toc,page,title]{appendix}
\usepackage{comment}
\usepackage{afterpage}
\usepackage{mathtools}
\usepackage{multirow}
\usepackage{array,booktabs}
\newcolumntype{M}[1]{>{\centering\arraybackslash}m{#1}}
\DeclareMathOperator{\diag}{diag}
\DeclareMathOperator{\dist}{dist}

\DeclareMathOperator{\poly}{poly}
\DeclarePairedDelimiter\ceil{\lceil}{\rceil}

\title{An Analytical Representation of the 2d Generalized Balanced Power Diagram}
\author{Christian Jung$^{1}$ \and Claudia Redenbach$^1$}

\date{
	{\small $^1$RPTU Kaiserslautern-Landau \\  Gottlieb-Daimler-Straße 48, 67663 Kaiserslautern, Germany}
}

\begin{document}

\maketitle

\textbf{Abtract}\\
Tessellations are an important tool to model the microstructure of cellular and polycrystal materials. Classical tessellation models include the Voronoi diagram and Laguerre tessellation whose cells are polyhedra. Due to the convexity of their cells, those models may be too restrictive to describe data that includes possibly anisotropic grains with curved boundaries. Several generalizations exist. The cells of the generalized balanced power diagram are induced by elliptic distances leading to more diverse structures. So far, methods for computing the generalized balanced power diagram are restricted to discretized versions in the form of label images. %Neither an analytic description of the diagram's geometry nor an explicit algorithm to compute it is available. 
In this work, we derive an analytic representation of the vertices and edges of the generalized balanced power diagram in 2d. Based on that, we propose a novel algorithm to compute the whole diagram.\\

\textbf{Keywords}\\
Voronoi diagram, Laguerre tessellation, anisotropy, non-convexity, curved boundary, conic sections

\section{Introduction}
Tessellations play an important role in many fields of research including computational geometry, computer graphics, materials science, and stochastic geometry. A tessellation can be interpreted as a subdivision of space into subsets, called cells. A well-studied class of tessellations are Voronoi-like tessellations which are generated by a finite set of points in $\mathbb{R}^d$. Every point in space is assigned to the generator it is closest to w.r.t. a given distance measure $d$. Depending on the choice of $d$, a wide range of tessellation models emerges. An overview over distance functions and tessellations that have been studied in this context is given in \cite{deza}.

In case of the Euclidean distance, the Voronoi diagram \cite{okabe} is obtained. Assigning a weight $w_i\in\mathbb{R}_{>0}$ to every generator, the power distance between two points $x,\,p_i\in\mathbb{R}^d$ is defined by $\dist(x,p_i,w_i)=||x-p_i||^2-w_i$. The tessellation model induced by the power distance is called power diagram or Laguerre tessellation. It can be seen as a first generalization of the Voronoi diagram with its weights allowing to control the size of the cells. The cells of both the Voronoi diagram and the Laguerre tessellation are convex polytopes with their boundaries being subsets of hyperplanes. Algorithms for computing these models are implemented in, for example, CGAL \cite{cgal}, voro\texttt{++} \cite{voropp} or Neper \cite{neper}.

Voronoi and Laguerre tessellations have been used to reconstruct and stochastically model 2d and 3d cellular materials such as polycrystalline structures in aluminum alloys or foams \cite{andre, claudia1, claudia2, claudia3, vannuland, optimic, lyckegaard}. Several algorithms exist to optimize the configuration of generators such that the resulting tessellation describes the given structure as good as possible with respect to some discrepancy measure. These optimization techniques include simulated annealing \cite{andre-stereo}, cross-entropy methods \cite{ulm-cross-entropy, ulm-cross-entropy-2} and the Metropolis-Hastings algorithm \cite{seitl-MH}. Furthermore, Laguerre tessellations of pre-defined cell areas can be generated \cite{givenVolumes}.

However, the cell convexity makes these models quite restrictive when it comes to approximating real-world microstructures which may exhibit curved cell boundaries and significant anisotropy. Non-convex cell shapes can be obtained with the Apollonius diagram (also additively-weighted Voronoi diagram or Johnson-Mehl tessellation). It is induced by an additively-weighted Euclidean distance. Similarly, the multiplicatively-weighted Voronoi diagram is induced by Euclidean distances that are weighted by a scalar factor. Both models have curved cell boundaries. Analytic descriptions of their cells' boundaries are known \cite{held, will} and algorithms for their simulation exist \cite{kara1,karaapo,kimkim3d,mwvd}.

The set Voronoi diagram \cite{schaller} generalizes the Apollonius diagram. Its cells are defined via the minimal Euclidean distances to arbitrarily-shaped objects. Analytic representations of the set Voronoi diagram are hard to derive and for most cases one has to rely on approximative algorithms. In \cite{pomelo}, geometric objects are sampled by a finite number of points from which the Voronoi diagram is computed. Cells whose generators belong to the same object are merged. This results in the edges being connected line segments.

Voxelized Voronoi-like tessellations have been used to reconstruct polycrystalline materials as well \cite{ebsd-ulm, linproGBPD}. The software DREAM.3D \cite{dream3d1,dream3d2} uses a region-growing algorithm to expand an initial set of generators, each given as a set of connected and labeled pixels, until each pixel is assigned a unique label. 

A tessellation model with elliptic distances has been introduced in \cite{LSorig} as the anisotropic Voronoi diagram. It may exhibit curved boundaries and it is possible for a cell to have only one or two neighboring cells. Furthermore, the cells do not need to be connected in which case the generator is absent in at least one part of the cell. These parts are called orphans. Conditions for the anisotropic Voronoi diagram to be orphan-free have been studied in \cite{orphanfree1} and it can be shown that their duals are triangulations \cite{orphanfree2}. First explorations for reconstructing microstructures with anisotropic Voronoi diagrams have been made in \cite{jeulin}.

The generalized balanced power diagram (GBPD) \cite{linproGBPD} extends the anisotropic Voronoi diagram by weighting its elliptic generators. Its generators can be interpreted as weighted ellipsoids. This leads to more diverse cell shapes and sizes compared to the Laguerre tessellation. For the approximation of polycrystals, the GBPD proves to be superior to models with less parameters such as the Laguerre tessellation \cite{datadriven}, but comes with a higher computational effort. The configuration of generators is usually estimated by means of principal component analysis \cite{linproGBPD} and the weights can be derived from the individual grain sizes \cite{tef}. Model evaluation and comparison is mostly based on discretized versions of the diagrams and is often carried out by counting pixel mismatches. To improve the goodness of initial approximations, several optimization techniques exist, including a linear integer programming approach \cite{linproGBPD}, simulated annealing \cite{ulm2} or gradient-descent-based methods \cite{ulm3}.

However, these applications of the GBPD are based on approximate computations of the GBPD. Most of them include discretized versions in the form of label images \cite{ulm1,ulm2,linproGBPD,grainmaps} or, for 2d, simulations via intersecting level sets \cite{vorosweep}. A characterization that could be used for constructing the GBPD has been suggested \cite{curvedVor}. However, it does not yield a description of the diagram's topology and geometry. To the best of our knowledge, explicit algorithms or implementations are not available.

In this work we develop an approach to compute the 2d GBPD analytically. We observe that the vertices are intersections of conics and that the edges of the GBPD can be described in terms of parametric equations. The advantage of our algorithm is that it is an explicit algorithm that is easy to implement. It includes a geometric characterization of the diagram's edges and vertices from which cell perimeter and cell area can be computed directly.

The paper is structured as follows. In Section \ref{sec:tess} we give basic definitions of a tessellation and introduce the GBPD. Then, we summarize its properties in Section \ref{sec:prop}. In Section \ref{sec:comp}, we first derive a parametric representation of the diagram's edges and describe how its vertices can be computed. Based on that, we sketch an algorithm for computing the whole GBPD. Furthermore, we provide formulas for computing the cells' perimeters and areas. Finally, Section \ref{sec:conc} draws a conclusion and gives an outlook to future research.

\section{Tessellations}\label{sec:tess}

Let $S$ be a metric space. A tessellation of $S$ is a countable collection of closed sets $T=\{C_i\subset S \mid i\in\mathbb{N}\}$ with 
	\begin{enumerate}
		\item $\bigcup_{i} C_i = S$,
		\item $\mathring{C_i}\cap\mathring{C_j} = \emptyset$ for all $i\neq j$ where $\mathring{C_i}$ denotes the interior of $C_i$,
		\item $\#\{C_i \in T \mid C_i\cap B \neq\emptyset\}<\infty$ for all bounded $B\subset S$
	\end{enumerate}

The closed sets $C_i$ are called cells of the tessellation. \\

Let $P=\{p_1, \ldots ,p_n\} \subset S$ with $2 \leq n < \infty$ and $\dist$ a (possibly weighted) distance measure on $S$. The cells given by 
\begin{equation*}\label{tessdist}
	C_i=\{x\in S \mid \dist(x,p_i) \leq \dist(x,p_j) \ \forall\ p_j\in P\}
\end{equation*}
define a tessellation of the space $S$. The point $p_i\in P$ is called seed or generator of the cell $C_i$ and $P$ is called the generator set of the tessellation.

The set of cells induced by the distance $$\dist(x,p_i) = \dist_{V}(x,p_i)\coloneqq\lVert x-p_i\rVert_2$$ define the Voronoi diagram. 

A generalization of the Voronoi diagram is the Laguerre tessellation. It is generated by spheres $S_1,\ldots,S_n$ and induced by the power distance. For a sphere $S_i$ centered at $p_i$ with radius $r_i$, the power distance is given by
$$\dist(x,p_i) = \dist_{L}(x,p_i,r_i)\coloneqq\lVert x-p_i\rVert_2^2 - r_i^2.$$ 
For each point $x$ outside a sphere $S(p_i,r_i)$, $\dist_L (x,p_i,r_i)$ is the squared length of the tangent line from $x$ to the sphere \cite{claudiaThesis}. The alternative form of the power distance as given in the introduction is obtained by setting $w_i=r_i^2$.

Another tessellation that generalizes the Voronoi diagram is the multiplicatively-weighted Voronoi diagram which is given by the cells induced by the distance $$\dist(x,p_i,\sigma_i) = \dist_{W}(x,p_i,\sigma_i)\coloneqq\frac{\lVert x-p_i\rVert_2}{\sigma_i}$$
with $\sigma_1,\ldots,\sigma_n\in\mathbb{R}_{>0}$.

A generalization to the tessellations above is given as follows. Let $M_1,\ldots,M_n$ be positive definite $d\times d$-matrices and $w_1,\ldots,w_n\in\mathbb{R}$. The set of cells induced by 
\begin{equation}\label{gbpd}
\dist(x,p_i)=\dist_G (x,p_i,M_i,w_i)\coloneqq (x-p_i)^\top M_i (x-p_i) - w_i
\end{equation}
is called the generalized balanced power diagram (GBPD). 

A positive definite matrix $M$ can be decomposed into
\begin{equation}\label{decompo}
    M= U\Lambda^{-1} U^\top
\end{equation}
where $U=(u_1,\ldots,u_d)$ is an orthogonal matrix and $\Lambda= \text{diag}(a_1,\ldots,a_d)$ is a diagonal matrix. Consider the equation
\begin{equation}\label{drawellipse}
(x-p)^\top M (x-p) - w = 1.
\end{equation}
For $y=(y_1,\ldots,y_d)^T\coloneqq U^T(x-p)$, Equation (\ref{drawellipse}) resolves to
\begin{equation}\label{drawellipse1}
\frac{y_1^2}{a_1}+\ldots+\frac{y_d^2}{a_d} - w = 1.
\end{equation}
For $w=0$, Equation \eqref{drawellipse1} is the general equation for an ellipsoid. Because $y$ is a (length-preserving) rotation or reflection of $(x-p)$, the pair $(p,M)$ can be interpreted as an ellipsoid centered at $p$ with semi-axes $u_1,\ldots,u_d$ and semi-axis lengths $\sqrt{a_1},\ldots,\sqrt{a_d}$. Thus, the generators of the GBPD can be seen as ellipsoids that expand non-linearly with a speed differing in each direction, which results in possibly anisotropic cell shapes.  

Note that adding the same constant to every weight does not change the diagram. Therefore, assume w.l.o.g. $-1<w$. Equation \eqref{drawellipse1} can then be re-written as\begin{equation}\label{drawellipse2}
\frac{y_1^2}{(1+w)a_1}+\ldots+\frac{y_d^2}{(1+w)a_d} = 1.
\end{equation}
Hence, $w$ can be interpreted as a non-linear scaling factor, scaling the ellipsoid equally in every direction by a factor of $\sqrt{1+w}$.

Note that the distance in Equation \eqref{gbpd} is not necessarily a metric since it can have negative values. 

The GBPD generalizes the Laguerre tessellation which is obtained for $M_i=I$ and $w_i=r_i^2$. It also generalizes the multiplicatively-weighted Voronoi diagram which we obtain for $w_i=0$ and $M_i = \nicefrac{1}{\sigma^2_i} I$.

\section{Properties of the GBPD}\label{sec:prop}

For the remainder of this paper, we focus on the case $d=2$. In this section, we visualize the GBPD's most important properties. A generator $i$ is visualized by all points inside the scaled elliptic contour given by Equation \eqref{drawellipse}. 

\begin{figure}[H]
	\centering
		\includegraphics[width=190px]{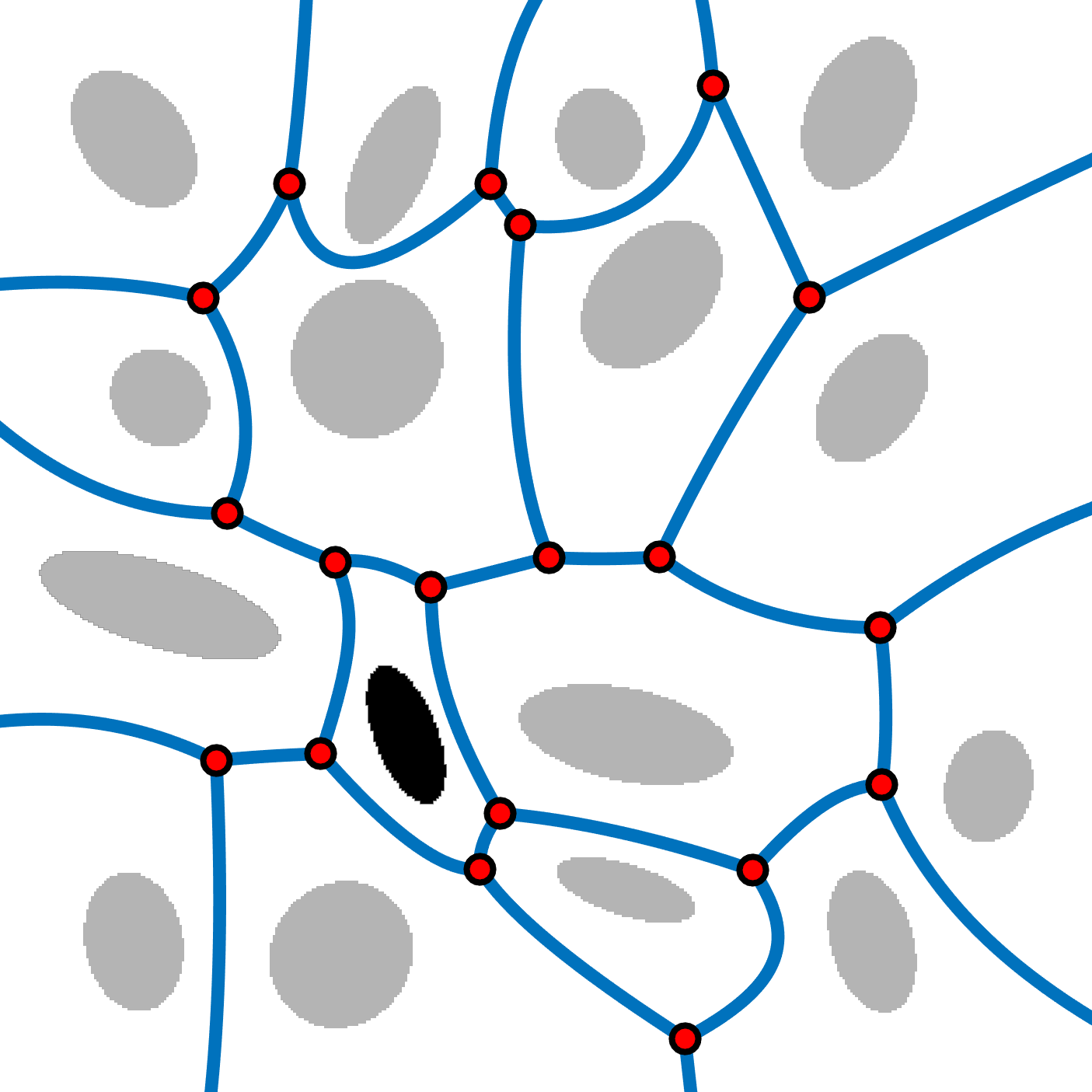}
		\includegraphics[width=190px]{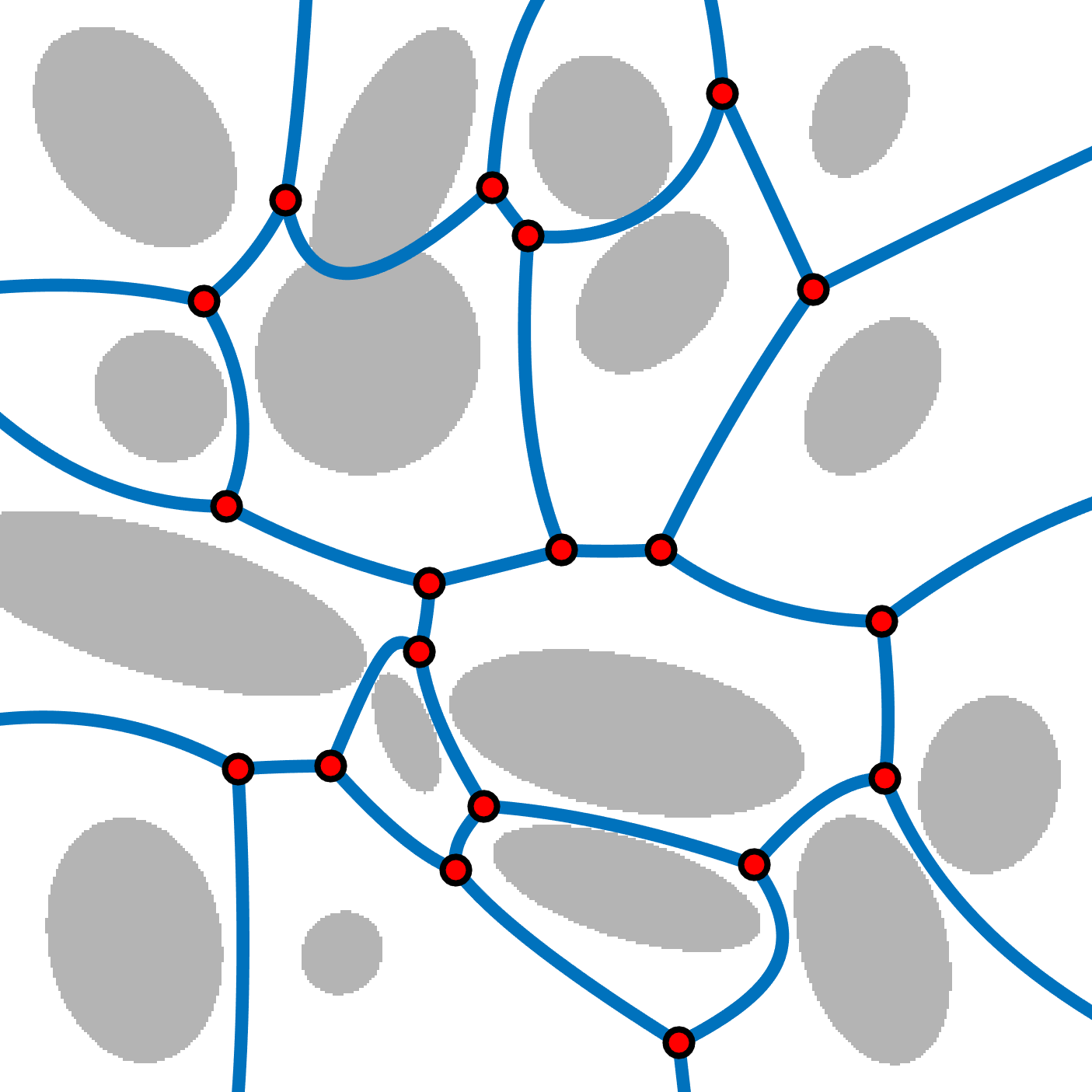}

	\caption{The GBPD generated by a set of 16 generators with $w_i=0$ (left) and $w_i$ drawn from a $(-1,3)$-uniform distribution (right).}
	\label{GBPD-visu1}
\end{figure}

Figure \ref{GBPD-visu1} shows the GBPD generated by 16 ellipses inside a $400\times 400$ window with different values of $w_i$. In the left image, $w_i=0$ for all $i$ while in the right image the $w_i$ are drawn from a uniform distribution on $(-1,3)$ resulting in the cells' areas becoming larger or smaller.

\begin{figure}[H]
	\centering
	\includegraphics[width=190px]{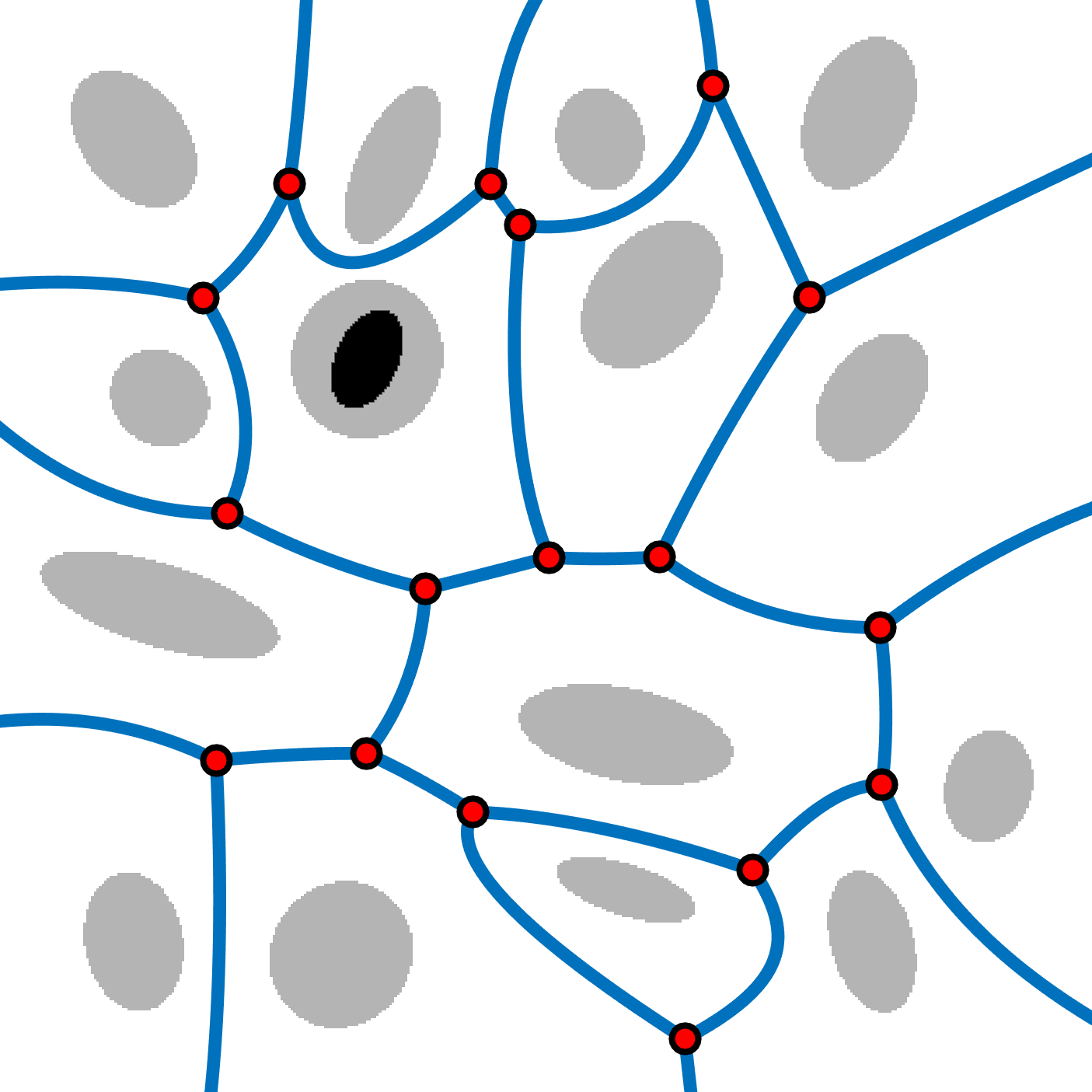}
	\includegraphics[width=190px]{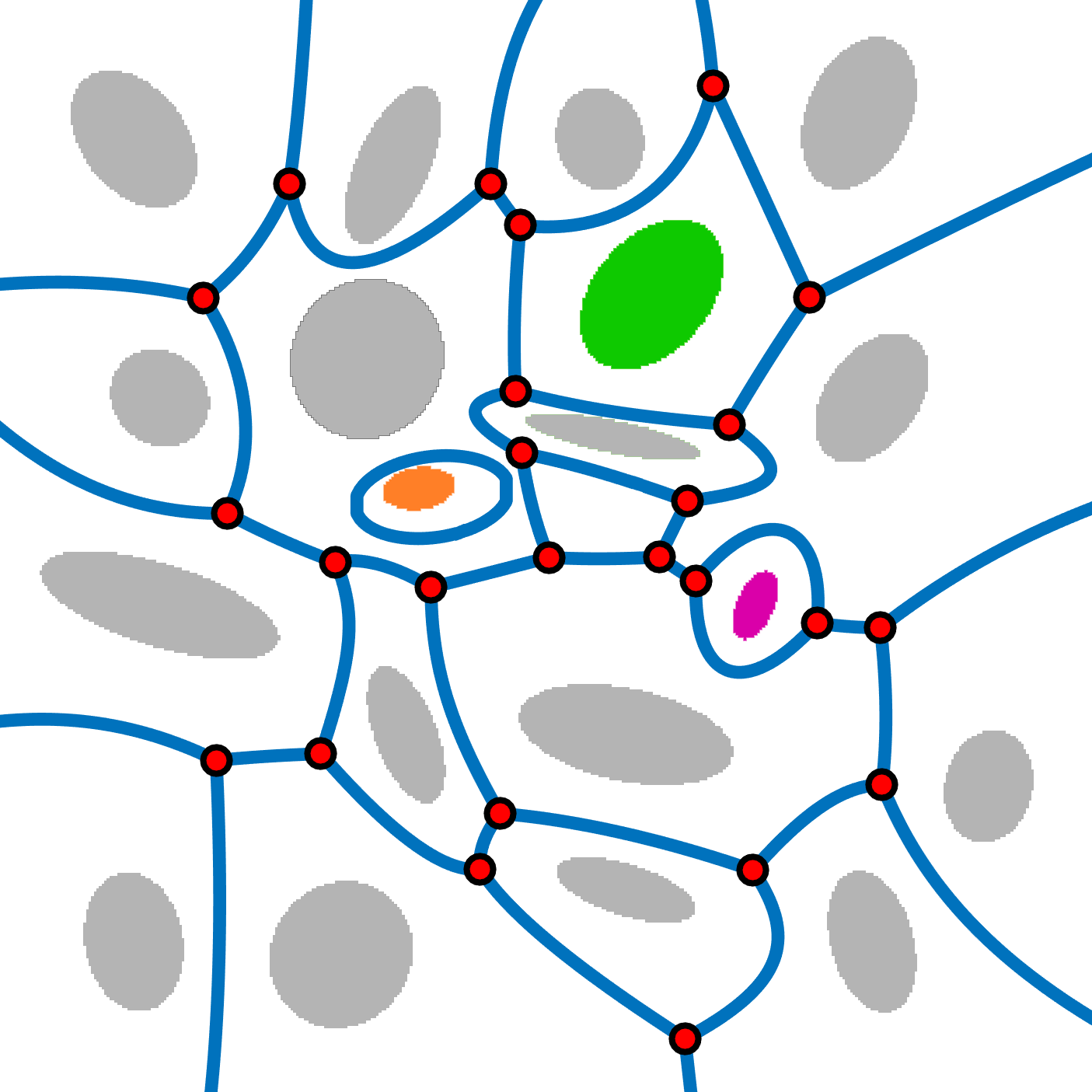}
	
	\caption{Left: The GBPD generated by a set of 17 generators with two of them generating empty cells. Right: The GBPD generated by a set of 19 generators. The cell of the orange generator has one neighbor, the cell of the pink generator has two neighbors and the green generator generates a cell that is disconnected. }
	\label{GBPD-visu2}
\end{figure}

For very small values (relative to the other weights), it is also possible that a generator has an empty cell which can be seen in the left image of Figure \ref{GBPD-visu2}. Here, the black generator from Figure \ref{GBPD-visu1}, left, is assigned a weight of $-100$. Empty cells may also be obtained for generators that lie inside another generator such as the black generator in Figure \ref{GBPD-visu2}, left.

We further note that the cells may have less than three neighbors. This is visualized in Figure \ref{GBPD-visu2}, right. Cells with exactly two neighbors are also called lenses. Furthermore, cells need not be connected.

\section{Computing the GBPD}\label{sec:comp}

As pointed out in \cite{curvedVor}, the GBPD in $\mathbb{R}^d$ is the restriction of the Laguerre tessellation in $\mathbb{R}^D$, $D=\nicefrac{d(d+3)}{2}$, to the $d$-manifold in $\mathbb{R}^D$ given by
$$\mathcal{Q}=\{\phi(x),\,x\in\mathbb{R}^d\}$$
with
$$\phi(x)=(x,\tilde\phi(x))$$
where
$$\tilde\phi(x)=(x_r\cdot x_s,\,1\leq r\leq s\leq d).$$
However, implementing this method for constructing the GBPD is not easy and to the best of our knowledge, an implementation is not available. Boissonnat et al. \cite{curvedVor} sketches an algorithm for computing the GBPD's vertices, but does not give an algorithm for computing the whole diagram.

In this section, we propose an algorithm that is only based on the definition of the GBPD. 

\subsection{Edges}\label{sec:edges}
The cells of a 2d Laguerre tessellation are convex polytopes with their edges being straight line segments. The cells of the GBPD are not necessarily convex as the cells' edges are generally curved. Via Equation \eqref{gbpd}, the bisector between two generators $i$ and $j$ can be expressed as the set of all $q=(x,y)^T\in\mathbb{R}^2$ satisfying

\begin{equation}
(q-p_i)^\top M_i (q-p_i) - w_i = (q-p_j)^\top M_j (q-p_j) - w_j
\end{equation}

which yields the implicit equation

\begin{equation}\label{implicit}
E_{ij}=(q-p_i)^\top M_i (q-p_i) - w_i - (q-p_j)^\top M_j (q-p_j) + w_j = 0.
\end{equation}

With 
\begin{align*}
\begin{pmatrix}
a_{11}	& a_{12} 	 \\
a_{21}  & a_{22} 	  
\end{pmatrix}=A_{ij} & \coloneqq M_i-M_j,\\
\begin{pmatrix}
b_{11}	& b_{12} 	 
\end{pmatrix}^T=B_{ij} & \coloneqq -2(M_i p_i-M_j p_j),\\
c & \coloneqq p_i^T M_i p_i - p_j^T M_j p_j - w_i + w_j
\end{align*}

and $A_{ij}$ being symmetric, Equation \eqref{implicit} is equivalent to 
\begin{equation}\label{conicImp}
	a_{11} x^2 + 2 a_{12} xy + a_{22} y^2 + b_{11} x + b_{12} y + c = 0.
\end{equation}
That is, the edges of the GBPD are parts of quadratic curves and, in particular, are conic sections. Our goal is to convert the conic's implicit form to a parametric representation of the form 
\begin{align*}
E:\mathbb{R}&\rightarrow \mathbb{R}^2\\
t&\mapsto (x(t),y(t)).
\end{align*}
Several conversion methods exist, including a method based on minimal $\mu$-bases \cite{mu-bases}, a pencil-of-lines method \cite{bookhoffmann, techrephoffmann} and a projective transformation method. Here, we use the latter. Details are elaborated in \cite{bookhoffmann} and \cite{techrephoffmann}.

The idea is to describe the conic in terms of homogeneous coordinates. Using translations and rotations, the conic is then transformed to a conic for which a parametric representation is known. Finally, the inverse transformations are applied to that representation.

Consider the homogeneous coordinate vector $z=(\tilde x,\tilde y,\tilde u)^T$ and

$$D=\begin{pmatrix}
a_{11}	& a_{12} & \nicefrac{b_{11}}{2}	 \\
a_{12}  & a_{22} & \nicefrac{b_{12}}{2} \\
\nicefrac{b_{11}}{2} & \nicefrac{b_{12}}{2} & c\end{pmatrix}.$$

For $\tilde u=1$, the homogeneous form of Equation \eqref{conicImp} is 

\begin{equation}\label{conicHomog}
z D z^T = 0.
\end{equation}

Since $D$ is symmetric, it can be diagonalized via $S=T^{-1}DT=\diag(\lambda_1,\lambda_2,\lambda_3)$. In case of at least one diagonal element of $S$ being equal to $0$, the conic consists of lines. For the more general case $\lambda_1,\lambda_2,\lambda_3\neq 0$, the conic's parametric form depends on the signs of the $\lambda_i$. If the signs of all $\lambda_i$ are equal, the conic is imaginary. Here, this case is irrelevant as the original conic is real-valued and remains real-valued after applying real-valued transformations. Setting $\mu_i = \nicefrac{1}{\sqrt{|\lambda_i|}}$, the parametric forms for the remaining cases are summarized in Table~\ref{table:param}. Computing $(\hat x(t),\hat y(t),\hat u(t))^T=T(\tilde x(t),\tilde y(t), \tilde u(t))^T$, a parametric form of the conic in Cartesian coordinates is given by 
\begin{align*}
E:\mathbb{R}&\rightarrow \mathbb{R}^2\\
t&\mapsto (x(t),y(t)).
\end{align*}
with
\begin{align}
\begin{split}
x(t) = \hat x(t) / \hat u(t),\\
y(t) = \hat y(t) / \hat u(t).
\end{split}
\label{param-cart}
\end{align}

\begin{table}[H]
\centering
	\begin{tabular}{ |c|c|c|c|c| } 
		\hline
		$\lambda_1$ & $\lambda_2$ & $\lambda_3$ & object & parametric form \\
		\hline
		 \multirow{4}{1em}{\centering + -} & \multirow{4}{1em}{\centering + -} & \multirow{4}{1em}{\centering - +} & \multirow{4}{*}{ellipse or circle} & \multirow{4}{10em}{\centering $\tilde x(t)=(1-t^2)\mu_1$ $\tilde y(t)=2t\mu_2$ $\tilde u(t)=(1+t^2)\mu_3$} \\ 
		  &  &  & &\\
		 & & & & \\
		 & & & & \\
		\hline
		\multirow{4}{1em}{\centering + -} & \multirow{4}{1em}{\centering - +} & \multirow{4}{1em}{\centering - +} & \multirow{4}{*}{hyperbola} & \multirow{4}{10em}{\centering $\tilde x(t)=(1+t^2)\mu_1$ $\tilde y(t)=2t\mu_2$ $\tilde u(t)=(1-t^2)\mu_3$} \\ 
		&  &  & &\\
		& & & & \\
		& & & & \\
		\hline
	\multirow{4}{1em}{\centering - +} & \multirow{4}{1em}{\centering + -} & \multirow{4}{1em}{\centering - +} & \multirow{4}{*}{hyperbola} & \multirow{4}{10em}{\centering $\tilde x(t)=2t\mu_1$, $\tilde y(t)=(1+t^2)\mu_1$, $\tilde u(t)=(1-t^2)\mu_3$} \\ 
	&  &  & &\\
	& & & & \\
	& & & & \\
		\hline
	\end{tabular}
\caption{Parametric forms of conics.}\label{table:param}
\end{table}

\subsection{Vertices}\label{sec:vertices}
The vertices of the GBPD are the locations that are equidistant to three or more generators. Practically, they can be found by intersecting two edges $E_{ij}$, $E_{jk}$ of the GBPD that are associated to a common generator. This can be realised via a pencil-of-conics method \cite{conicIntersect} which is already implemented as a MATLAB package \cite{conicIntersectMatlab}.

Intersecting two conics may result in up to four intersecting points. Not all of these points are part of the GBPD as they may belong to the cell of another generator. Which vertices are part of the diagram can be checked by computing the distances to every generator.

Let $v=(v_1,v_2)\in\mathbb{R}^2$ be an intersection point of the curves $E_{ij}$ and $E_{jk}$ that is part of the GBPD. The parameter $t\in\mathbb{R}$ in Equation \eqref{param-cart} for $E_{ij}$ (or $E_{jk}$) corresponding to $v$ can then be retrieved by solving the quadratic equations $(v_1,v_2)=(x(t), y(t))$ for $t$.

\subsection{An algorithm for simulating the GBPD}\label{sec:alg}

With the analytic representations of the edges and vertices of the GBPD we propose the following algorithm for computing the GBPD:

\begin{enumerate}
	\item Input: Set of generators $G=(p_i,M_i,w_i)$.
	\item For every triple of generators compute the intersection points of their edges. For each intersection point, check whether it is a vertex of the diagram.
	\item For every triple that yields at least one GBPD vertex, create a list for each of the pairs $(i,j), (j,k), (i,k)$ and store the vertices. If the list already exists, append.
	\item For each list draw the visible parts of the corresponding bisectors: Solve the parametric form of a bisector $E_{ij}$ (cf. Equation (\ref{param-cart})) for $t$ for each vertex on that bisector. Sort the values in ascending order resulting in  $t_1<t_2<\ldots<t_{l-1}<t_l$. The edge from $t_i$ to $t_{i+1}$ is part of the diagram if $(x(\hat t),  y(\hat t))$ for  $\hat t = \nicefrac{(t_i + t_{i+1})}{2}$ is closest to $i$ and $j$ among all generators (since otherwise another vertex would exist). Additionally, check if the edges from $-\infty$ to $t_1$ and $t_l$ to $\infty$ are part of the diagram.
	\item For each pair of generators that is not part of a vertex-generating triple, check whether the whole bisector is part of the diagram by checking if $(x(0), y(0))$ is closest to $i$ and $j$ among all generators.
	\item Output: Edges and vertices of the GBPD including adjacencies
\end{enumerate}

For a run-time analysis of the algorithm first consider a set of $n$ generators that produces a GBPD without disconnected cells and with each cell having least three neighbors. Such a tessellation can be considered a planar graph. Let $|m|$, $|e|$, $|v|$ be the number of its cells, edges and vertices. Since we assume only connected cells, it holds that $|m| \leq n$ and thus $|m|$ lies in $\mathcal{O}(n)$. Euler's formula for planar graphs also implies that $|e|$ and $|v|$ each lie in $\mathcal{O}(n)$.

In step 2 of the algorithm, $\binom{n}{3}$ intersections are computed and for each the distance to all $n$ generators is determined. An edge intersection is computed in time that is constant with respect to the input size $n$. In total, this results in $\mathcal{O}(\binom{n}{3}\cdot n)$ time.

In step 4, $\mathcal{O}(n)$ edge segments are considered and for each a distance to $n$ generators is computed. A bisector computation runs in time that is constant with respect to the input size $n$. In total, step 4 results in a time of order $\mathcal{O}(n^2)$.

Lenses consist of only two vertices and edges, and thus are less complex than regular cells. Hence, the linear relation between the number of faces, edges and vertices is preserved if we allow the GBPD to have lenses. This also holds for cells that have only one neighbor since they only consist of one edge and zero vertices. In this case, step 5 of the algorithm needs to be carried out as well, resulting in an additional time of order $\mathcal{O}(\binom{n}{2}\cdot n)$. This yields a total run-time for the algorithm of order $\mathcal{O}(\binom{n}{3}\cdot n)$.

We have restricted our run-time analysis to connected cells such that the number of connected components is less or equal than the number of generators. To extend the analysis to disconnected cells a better understanding of the relation between $n$ and the number of connected components is necessary.

\begin{figure}[H]
	\centering
	\includegraphics[width=120px]{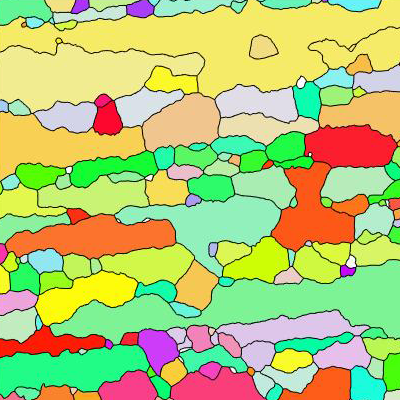}
	\includegraphics[width=120px]{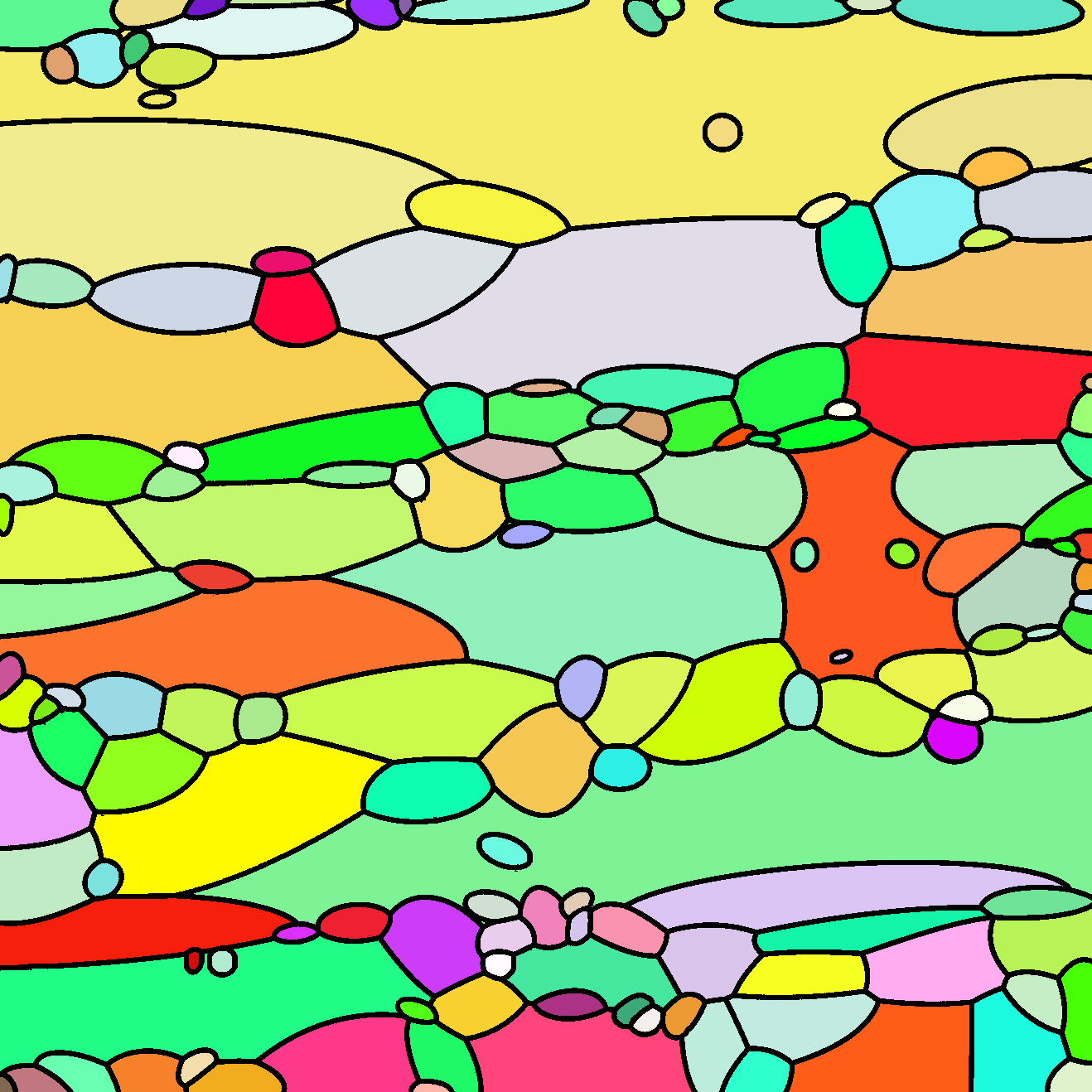}
	\includegraphics[width=120px]{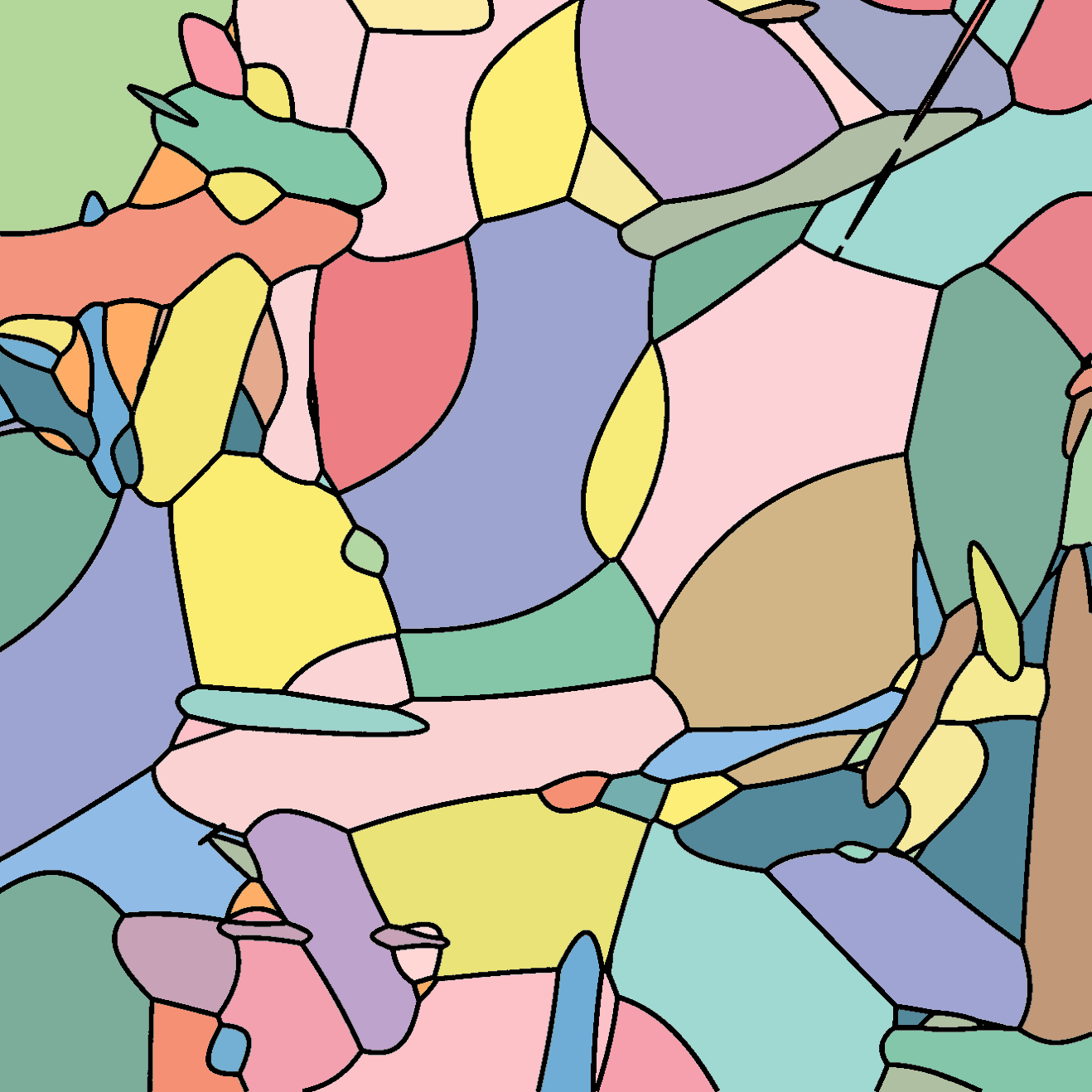}
	
	\caption{Left: 2d EBSD image from which 165 generators are extracted via PCA. Middle: The resulting GBPD. Right: The GBPD generated by a set of 148 uniformly distributed generators. }
	\label{GBPD-alg1}
\end{figure}

For validation of the algorithm, we present two examples. Figure \ref{GBPD-alg1}, left, shows an EBSD image of size $400\times 400$ and Figure \ref{GBPD-alg1}, middle, shows the reconstruction with the GBPD of 165 generators. They were extracted from the polycrystals of the EBSD image via principal component analysis (PCA) as described in \cite{linproGBPD}. The $w_i$ are all set to $0$. The diagram's edges that are output by our algorithm are shown in black. For comparison, the diagram is also computed pixel-wise in a ``brute-force'' way, assigning every pixel in the image to the generator it is closest to. This results in a label image. Its colors are derived from the original EBSD image.

The GBPD in Figure \ref{GBPD-alg1}, right, is generated by a set of 148 generators with centers $p_i=(p_{i,1}\; p_{i,2})^T$, $p_{i,1}$ and $p_{i,2}$ being uniformly distributed on $[0,400]$ and $w_i$ on $[0,50]$. The $M_i$ are constructed via Equation \eqref{decompo} 
with orthogonal matrices $U_i$ defined by 
$$U_i=\begin{pmatrix}
\cos\theta_i & -\sin\theta_i \\
\sin\theta_i & \cos\theta_i \end{pmatrix}$$
and $\theta_i$ being uniformly distributed on $[0,\pi]$. The major semi-axes lengths $\sqrt{a_{i,1}}$ are drawn from a uniform distribution on $[10,20]$ and the minor semi-axes lengths $\sqrt{a_{i,2}}$ from a uniform distribution on $[0,10]$

\subsection{Cell Perimeter and Area}\label{sec:area}

With the algorithm explained above it is possible to analytically compute cell perimeters and areas as well. The perimeter of a cell consisting of $h$ edges can be computed as

$$\mathlarger{\mathlarger{\mathlarger{\sum}}}_{i=1}^h \mathlarger{\mathlarger{\int}}_{t_1^i}^{t_2^i} \sqrt{\left(\frac{\partial }{\partial t} x(t) \right)^2 + \left(\frac{\partial}{\partial t} y(t) \right)^2}dt$$

where $t_1^i$ and $t_2^i$ are the parameters corresponding to the start point $v_1^i$ and end point $v_2^i$ of an edge $e_i$.

To compute the area of a cell, consider a connected cell. Let $\poly(V)$ be the polygon that is obtained by traversing the cell's vertices $V$ in clockwise or counter-clockwise order and $|\poly(V)|$ be its area. First, compute $|\poly(V)|$. Then, compute the area between $e_i$ and the straight line from $v_1^i$ to $v_2^i$ and add it to the total area if $e_i$ lies outside the polygon. Otherwise, subtract. In practice, this can be done by translating and rotating the line from $v_1^i$ to $v_2^i$ such that $v_1^i=(0,0)$ and the line lies on the $x$-axis. The same transformations are applied to $e_i$. Then the area between the line and $e_i$ is given as 

$$A_{i}=\mathlarger{\mathlarger{\int}}_{t_1^i}^{t_2^i} \left|\bar y(t) \left(\frac{\partial}{\partial t} \bar x(t) \right)\right| dt$$

where $(\bar x(t),\bar y(t))$ is the parametric representation of the rotated edge.

This yields

$$A=|\poly(V)|-\sum_{i=1}^n A_i \mathds{1}_{\poly(V)}(e_i) + \sum_{i=1}^n A_i \mathds{1}_{(\poly(V))^C}(e_i)$$

for the total cell area.

In case of a cell with a single neighbor, the area can be computed directly by computing the area of an ellipse or circle. It is subtracted from the area of the cell it is enclosed in.
In case of a lense, only two integrals need to be summed up.
In case of disconnected cells, the approach above needs to be executed for each part of the cell separately.

\section{Conclusion}\label{sec:conc}

In this work we have derived a representation of the edges of the GBPD in parametric form. Its vertices correspond to intersections of conics. With that we have formulated a novel and explicit algorithm that computes the GBPD. Analytic formulas for computing cell perimeter and area can be applied and have been implemented as well. Furthermore, topological and adjacency information of the diagram can be stored during the algorithm. 

From a computational point of view, more efficient implementations of the algorithm are necessary. As discussed in Section \ref{sec:comp}, the $d$-dimensional GBPD relates to the $D$-dimensional Laguerre tessellation where $D=\nicefrac{d(d+3)}{2}$. The latter can be constructed in $\mathcal{O}(n^{\ceil*{\frac{D}{2}}})$ time \cite{aurenhammer} which suggests that the run-time of our algorithm is not optimal. However, to the best of our knowledge, our algorithm is the first algorithm to compute the 2d GBPD explicitly.

For future work, the algorithm may be applied in the context of inverse problems such as reconstructing cellular and polycrystalline materials. The analytic, geometric descriptions can be used to define further goodness measures to evaluate the reconstruction of these materials. 

In addition, we focus on the extension of the algorithm to 3d. Then, the algorithm may be used to reconstruct 3d materials such as foams or 3d EBSD images.

\section*{Acknowledgement}
This work was supported by the German Federal Ministry of Education and Research (BMBF) [grant number 05M2020 (DAnoBi)].

\bibliography{main}
\bibliographystyle{plain}
\end{document}